\documentclass[traditabstract,rnote]{aa}
\usepackage{natbib} 
\usepackage{graphicx}
\usepackage{txfonts}

\begin{document}

\title{Are 3C\,249.1 and 3C\,334 restarted quasars?}

\author{Andrzej Marecki} \offprints{Andrzej Marecki 
\email{amr@astro.uni.torun.pl}}

\institute{Toru\'n Centre for Astronomy, N. Copernicus University,
           87-100 Toru\'n, Poland}

\date{Received 13 July 2012 / Accepted 23 August 2012}

\abstract {This {\em Research Note} follows up a Letter in which I~posit 
that J1211+743 is a restarted radio source. This means that its structure, 
where the jet points to the relic lobe, is only apparently paradoxical. 
Here, I propose the same scenario and apply the same mathematical model to 
3C\,249.1 and 3C\,334. The ultimate result of my investigation is that these 
two well-known radio-loud quasars can be understood best so far if it was 
assumed that they, too, had been restarted.}

\keywords{Radio continuum: galaxies, Galaxies: active}

\maketitle


\section{Introduction}

In principle, the ``twin exhaust'' model of a classical radio source 
\citep{BR1974} predicts full symmetry of its doubled structures, i.e. lobes and 
jets, but this is very often not the case in reality. The main cause of 
asymmetries is the source orientation leading to beaming and 
Doppler-boosting that explain the apparent one-sidedness of the jets very 
well. However, asymmetries of the lobes cannot be attributed to relativistic 
beaming because they are not moving with relativistic velocities 
\citep{LR1979}. A good example of a class of object where asymmetries are 
conspicuous, but do not result from the orientation, are the so-called 
HYbrid MOrphology Radio Sources (HYMORS) \citep{GKW2000,Gawr2006}, i.e. 
sources each of whose sides are a different Fanaroff-Riley (FR) type 
\citep{FR1974}. Earlier, \citet{GWH1996} highlighted several radio-loud 
quasars in which the hotspot on the jetted side was barely detected. The 
authors coined the term ``weak-headed quasars'' for them and argued that the 
faintness of the nearer hotspot, i.e. the one on the jet side, stemmed from 
the differential light-travel time caused by the source orientation so that 
the near-side lobe was viewed at a later evolutionary stage as compared to 
the far-side lobe.

\citet{Mar2012} -- hereafter Paper~I -- independently used the same argument 
to show that in \object{J1211+743} (4CT\,74.17.01), a source which, at first 
sight, could be labelled a HYMORS, the asymmetric appearance of the lobes 
also resulted from differential light-travel time. Therefore, J1211+743 is a 
pseudo-HYMORS, not a real one, whereas the correlation between the jet 
direction and the location of the relic lobe is only apparently paradoxical. 
The case of J1211+743 is important for yet another reason. It has been shown 
in Paper~I that this object must have been restarted to appear the way it 
does. Consequently, not only the so-called double-double sources are those 
where activity restart surely takes place -- see the review of \citet{SJ2009}.
It thus seems timely to search for other objects similar to J1211+743 and, 
if found, to check using the formalism developed in Paper~I whether their 
observed geometry fits the activity re-ignition scenario combined with the 
differential light-travel time.

\section{In quest of J1211+743 analogues}

In the 607-MHz map of J1211+743 \citep{Pir2011}, the tip of its one-sided 
jet is located close to the boundaries of its relic lobe. However, from the 
point of view of the scenario presented in Paper~I, this circumstance is not 
at all essential for the model to remain valid. The exact location of the 
tip of the jet can be largely arbitrary within the lobe itself or its 
vicinity, although it obviously cannot coincide with the expected location of 
the hotspot, otherwise it would be hardly possible to prove that the relic 
had no hotspot. When looking for other objects analogous to J1211+743, therefore,
the candidates only have to be tested against the following two criteria:

\begin{enumerate}

\item The source has one FR\,II-like lobe with a hotspot and one diffuse 
relic lobe without a hotspot.

\item The source has a one-sided jet and the relic is on the jet side.

\end{enumerate}

I~visually inspected the images of double-lobed radio sources published in 
\citet{Bla1992}, \citet{Lon1993}, \citet{Pri1993}, \citet{Bog1994}, 
\citet{Bri1994}, \citet{Law1995}, \citet{Nef1995}, \citet{Rei1995,Rei1999}, 
\citet{Fer1997}, \citet{Har1997}, \citet{HH1998},  \citet{Ril1999},
\citet[][hereafter G04]{Gil2004}, \citet{Fer2007}, and \citet{Kha2008}
to find objects meeting these criteria. It turned out that only two sources 
-- \object{3C\,249.1} and \object{3C\,334} -- clearly met them, so they are 
discussed in detail below. Additionally, I~made an exception for 
\object{1317+520} and \object{2209+152}, and despite not meeting the 
criteria by these objects, I~consider them in Sect.\,2.4.

\subsection{The case of 3C\,249.1}

The object 3C\,249.1 (PG\,1100+772) is a $z=0.3115$ quasar whose radio
structure is best shown in Figs.\,5 and 6 of \citet{LM1983}, Figs.\,18--21
of \citet{Bri1994}, and Figs.\,37--39 of G04. The western lobe is FR\,II
type while the 
eastern one is devoid of a hotspot and remarkably featureless. This, 
together with a high value of the spectral index $\alpha \sim 0.9$ (G04), 
which is directly responsible for a complete absence of eastern lobe in the 
8.4-GHz image of 3C\,249.1 \citep{Fer2007}, combined with the diffuseness of 
that lobe, which is best seen in the 408-MHz image \citep{LM1983}, provide
compelling evidence that the eastern lobe is a relic.

A number of explanations of the peculiar morphology of 3C\,249.1 have been 
proposed in the literature -- see G04 for references -- including the 
``flip-flop'' mechanism \citep{LM1983}. According to these authors, the jet 
was once directed to the east, and it inflated the eastern lobe. After a 
reversal, which is the essence of the ``flip-flop'' model, the jet was 
oriented westerly so that the western lobe developed. The subsequent 
reversal is reflected in the direction of the jet as currently perceived. 
Although the ``flip-flop'' mechanism is no longer used to interpret 
double-lobed radio sources, one element of that theory still remains viable: 
the jet is currently not energising the eastern lobe because these two 
components belong to different epochs and their alignment is a mere spatial 
coincidence. This assertion is a cornerstone of the scenario I~propose for 
3C\,249.1, the same as the one devised for J1211+743 in Paper~I.

My model posits that, instead of the reversals of the {\em intrinsically} 
one-sided jet, the activity of the nucleus is recurrent and both jets were 
not produced for some time -- a quiescence period, $t_q$. The transition of 
the nucleus to the quiescent mode combined with a short lifetime of the 
hotspot \citep[$7\times 10^4$\,yr,][]{KSR2000} led to the swift disappearance 
of hotspots. However, if the source does not lie on the sky plane then the 
differential light-travel time makes it possible that for a limited period 
of time the relic is seen only on the near side. If the activity of the 
nucleus is reinstated within that time window then we observe a superposition
of two effects: the lobes' decay asymmetry and the presence of {\em apparently} 
one-sided jet pointing to the relic lobe. It has been shown in Paper~I that 
the above scenario requires fine-tuned timing, and the necessary formalism 
has been developed there. I~apply it here to test whether the model is valid 
for 3C\,249.1. The results are given in Sect.\,2.3.

\subsection{The case of 3C\,334}

The object 3C\,334 is a $z=0.5551$ quasar whose radio structure is best shown in 
Figs.\,24--26 of \citet{Bri1994}, and Figs.\,52--54 of G04. The western lobe 
is FR\,II type while the eastern one is somewhat peculiar. In the 4.9-GHz 
VLA map \citep[][Fig.\,25a]{Bri1994}, this lobe, although quite diffuse in 
general, has three features denoted as ``Q'', ``R'', and ``S''. None 
of these three dominates to make it a good candidate for the 
actual hotspot. On the other hand, in the 5-GHz VLA image shown in Fig.\,54 
in G04, the eastern lobe is rather featureless. G04 claim that there is a 
``clear trail of emission'' made of components ``S4'' and ``S5'' that 
provides a connection from feature ``S3'' in the jet all the way to feature 
``S6'' in the lobe. It could be, however, asserted equally well that there 
is a clear {\em gap} in emission between components ``S4'' and ``S5'' -- see 
the lower panel of Fig.\,54 -- and the lobe itself is isolated. Therefore, 
this image provides substantial ground for suspecting that the eastern lobe of 
3C\,334 is in fact a relic that is currently not fuelled by the jet that 
terminates with component ``S3'' at the lobe's outskirts.

I~thus propose the following interpretation of 3C\,334. The activity of its 
nucleus ceased at some point in the past. This led to the cut-off of the 
stream of relativistic plasma to both lobes we still observe. The western 
lobe is farther from us. That's why we perceive it as a being of FR\,II 
type, while the eastern lobe is seen as a relic of a former FR\,II-type 
lobe. After a period of quiescence $t_q$, the activity was renewed and the 
jet is a conspicuous signature of that. However, the central engine of 
3C\,334 was repositioned during the restart so now the jet misses the 
eastern lobe slightly and is not connected to it; the alleged bridge made of 
components ``S4'' and ``S5'' is an illusion caused by the bend of the jet.
(The bend of the jet is likely to be forced by the pressure gradient on
the boundary of the cocoon of the radio source.)

Based on the above scenario, application of the model developed in Paper~I 
to 3C\,334 seems to be justified. However, because of the misalignment 
between the straight section of the jet and the line connecting the core and 
the position of the tip of the relic lobe, the meaning of $d$ (see Fig.\,1 
of Paper~I) has changed -- it is now the length of the projection of the jet 
onto that line. Consequently, $\beta_{adv}$ is not the actual jet advance 
speed but the speed of the tip of the jet projection, so the true 
$\beta_{adv}$ is somewhat higher than the values of projected $\beta_{adv}$ 
used in the model.

\begin{table}[t]
\caption{Upper limits to the length of $t_q$ in 3C\,249.1 and 3C\,334 [Myr]} 
\begin{tabular}{c|c c c c|c c c c}
\hline
\hline
         & \multicolumn{4}{c|}{3C\,249.1} & \multicolumn{4}{c}{3C\,334}\\
           \cline{2-9}
$\theta$ & \multicolumn{4}{c|}{$\beta_{adv}$}
         & \multicolumn{4}{c}{$\beta_{adv}$}\\
$[\degr]$ & 0.2 & 0.25 & 0.3 & 0.35 & 0.2 & 0.25 & 0.3 & 0.35 \\
\hline
10 & 0.60 & 1.37 & 1.88 & 2.25 & 0.65 & 2.64 & 3.97 & 4.92 \\
20 & 0.25 & 0.64 & 0.90 & 1.09 & 0.20 & 1.21 & 1.88 & 2.36 \\
30 & 0.11 & 0.38 & 0.55 & 0.68 & ---  & 0.68 & 1.14 & 1.47 \\
40 & 0.02 & 0.23 & 0.37 & 0.46 & ---  & 0.37 & 0.73 & 0.98 \\
50 & ---  & 0.12 & 0.24 & 0.32 & ---  & 0.14 & 0.45 & 0.66 \\
60 & ---  & 0.04 & 0.14 & 0.22 & ---  & ---  & 0.23 & 0.42 \\
70 & ---  & ---  & 0.06 & 0.13 & ---  & ---  & 0.04 & 0.21 \\
80 & ---  & ---  & ---  & 0.05 & ---  & ---  & ---  & 0.03 \\
\hline
\hline
\end{tabular}
\end{table}

\subsection{Application of the model to 3C\,249.1 and 3C\,334}

In Paper~I, the constraints on the length of the quiescence period $t_q$ 
were analysed, and it was shown that $t_q$ must stay within certain limits 
for the model to be valid. To calculate the lower and the upper limits to 
$t_q$, one has to use Eqs.\,(2) and\,(4) (Paper~I), respectively. They 
require several parameters: $l_1$, $l_2$, $d$ -- see Fig.\,1 of Paper~I for 
the explanation of their meaning -- have to be extracted from the 
images of the sources; $\beta_{adv}$ and $\beta_{jet}$, which are common for 
3C\,249.1 and 3C\,334, are taken from the literature; and $\theta$ (see 
Fig.\,1 of Paper~I) is a free parameter. \citet{AL2004} calculated jet 
speeds for a number of double sources including 3C\,249.1 and 3C\,334 and 
they obtained $\beta_{jet}\approx 0.8$ for them. It is adopted here. As for 
$\beta_{adv}$, I use four values from the range $0.2 \leq \beta_{adv} \leq 
0.35$ that is similar to the one found for B\,1834+620 by \citet{Sch2000}.

On the basis of the map of 3C\,249.1 in G04 and using the standard 
cosmological parameters, I~adopted the following parameters for this source: 
$l_1\sin\theta=88$\,kpc, $l_2\sin\theta=128$\,kpc, $d\sin\theta=41$\,kpc. 
Since one of the lobes is diffuse, estimating the size of the whole
source reliably is hardly possible. What I did was simply to measure it
based on the range of the lowest contour in the published figure.
After substituting them to Eq.\,(2) of Paper~I, I~found that it was 
fulfilled for any value of $\theta$ if $\beta_{adv}=0.2$, for 
$\theta<78\fdg8$ if $\beta_{adv}=0.25$, for $\theta<61\fdg3$ if 
$\beta_{adv}=0.3$, and for $\theta < 47\fdg8$ if $\beta_{adv}=0.35$. Given 
that 3C\,249.1 is a quasar and that according to \citet{Bar1989} $\theta < 
44\fdg4$ for quasars, the above constraints on $\theta$ are not critical.
 
On the basis of the map of 3C\,334 in G04 and using the standard 
cosmological parameters, I~adopted the following parameters for this source: 
$l_1\sin\theta=206$\,kpc, $l_2\sin\theta=156$\,kpc, $d\sin\theta=106$\,kpc. 
As in the case of 3C\,249.1, I measured the source's size based on the 
range of the lowest contour in the published figure.
After substituting them to Eq.\,(2) of Paper~I, I~found that it was 
fulfilled for any value of $\theta$ if $0.2 \leq \beta_{adv} \leq 0.35$.

Substitution of the parameters for 3C\,249.1 and 3C\,334 to Eq.\,(4) of 
Paper~I yielded the upper limits to the length of the quiescent period for the
respective sources. They are given in Table\,1 for a wide range of values of 
$\theta$. The lack of a number at the crossing of a given row and column of 
Table\,1 means that the respective combination of parameters is not allowed 
if the scenario is to be viable. On the other hand, following the argument 
in Paper~I, the existence of upper limits for the remaining combinations of 
parameters proves that the model is plausible.

\subsection{The cases of 2209+152 and 1317+520}

The object 2209+152 is a $z=1.502$ quasar whose entire radio structure is 
best shown in Fig.\,159 of \citet{Lon1993}. It consists of a core, a curved 
jet and two lobes, both relics. It can be speculated that 2209+152 is 
restarted, i.e. the lobes were inflated in the previous active period but 
now are fading out due to the lack of fuelling, while the jet is a signature 
of the current active phase. The radio source 2209+152 is not featured by 
the lobe asymmetry that
could give us a clue to its orientation caused by differential light-travel 
time. Nevertheless, owing to the presence of a conspicuous Laing-Garrington 
effect \citep{Lai1988,Gar1988} in this source \citep[][Fig.\,A25]{Gar1991} 
it is confirmed that the source does not lie in the sky plane, the eastern, 
i.e. the jet-side, lobe being nearer the observer. It follows that 2209+152 
could possibly meet the criteria listed at the beginning of this section. An 
explanation for why it does not meet them anyway is quite straightforward: the 
actual quiescence period of its nucleus must have exceeded the upper limit 
calculated from the model for given parameters. As a result, the information 
that the far-side lobe had started to decay reached the location of the 
observer too early relative to the development of the jet and now we 
perceive that lobe as a relic. In other words, wavefront ``1'' passed the 
observer earlier than wavefront ``5'' -- see Fig.\,1 in Paper~I.

The above speculation can be extended. If the length of the quiescence 
period goes beyond the upper limit given by the model even further than in 
the case of 2209+152, then the near-side lobe vanishes, while the far-side 
lobe remains barely observable. A $z=1.061$ quasar 1317+520 imaged at 
several frequencies by \citet[][Fig.\,28]{Rei1995}, but see also 
\citet[][Fig.\,2]{JM2008}, is perhaps a good specimen to illustrate such a 
situation: there is no trace of the lobe on the jetted side, whereas the 
lobe on the opposite side is clearly a relic, because it is devoid of a 
hotspot and very diffuse.

\section{Concluding remarks}

Cessation of activity in galactic nuclei, possibly followed by its restart, 
is a phenomenon that can manifest itself in a number of ways, the existence 
of double-double radio sources being perhaps the most spectacular as shown
in the review of \citet{SJ2009}. A restarted radio source need not be 
symmetric, though. For example, \citet{MS2011} highlighted a small group of
galaxies where the production of the jet had stopped, leading to the decay of 
the radio lobes. Owing to the light-travel time difference between the 
lobes, that decay is perceived as asymmetric. In Paper~I, the effect of 
differential light-travel time was analysed quantitatively also taking the
activity re-ignition into account. The analysis carried out there proved that 
the scenario suggested for the sources with asymmetric lobes and the jet 
pointing to the relic was plausible for at least one source: J1211+743.

Here, I show that two well-known quasars, 3C\,249.1 and 3C\,334, can also be 
interpreted this way and that they have been restarted as well. The key 
parameter of my model is the upper limit to duration of the quiescent period 
between the two active periods, $t_q$. Possible values of the upper limits 
to $t_q$ for 3C\,249.1 and 3C\,334 have been displayed in Table\,1. They are 
lower that their counterparts for J1211+743 shown in Table\,1 of Paper~I. 
This can be attributed to smaller sizes of 3C\,249.1 and 3C\,334 as compared 
to J1211+743. Also, it is interesting to note that all the values of the 
upper limits to $t_q$ shown here are of the same order of magnitude as those
obtained for 
J0041+3224 and J1835+6204 by \citet{Kon2012} using a completely different 
approach.

\begin{acknowledgements}

\item This research has made use of the NASA/IPAC Extragalactic Database 
(NED), which is operated by the Jet Propulsion Laboratory, California 
Institute of Technology, under contract with the National Aeronautics and 
Space Administration.

\end{acknowledgements}


\begin{thebibliography}{}

\bibitem[Arshakian \& Longair(2004)]{AL2004}
Arshakian, T.~G., \& Longair, M.~S.\ 2004, \mnras, 351, 727 

\bibitem[Barthel(1989)]{Bar1989} Barthel, P.~D.\ 1989, \apj, 336, 606

\bibitem[Black et al.(1992)]{Bla1992} Black, A.~R.~S., Baum, S.~A.,
Leahy, J.~P., et al.\ 1992, \mnras, 256, 186

\bibitem[Blandford \& Rees(1974)]{BR1974} Blandford, R.~D.,
\& Rees, M.~J.\ 1974, \mnras, 169, 395

\bibitem[Bogers et al.(1994)]{Bog1994} Bogers, W.~J., Hes, R.,
Barthel, P.~D., \& Zensus, J.~A.\ 1994, \aaps, 105, 91

\bibitem[Bridle et al.(1994)]{Bri1994} Bridle, A.~H., Hough, D.~H., 
Lonsdale, C.~J., Burns, J.~O., \& Laing, R.~A.\ 1994, \aj, 108, 766

\bibitem[Fanaroff \& Riley(1974)]{FR1974} Fanaroff, B.~L.,
\& Riley, J.~M.\ 1974, \mnras, 167, 31P

\bibitem[Fernini et al.(1997)]{Fer1997} Fernini, I.,
Burns, J.~O., \& Perley, R.~A.\ 1997, \aj, 114, 2292

\bibitem[Fernini(2007)]{Fer2007} Fernini, I.\ 2007, \aj, 134, 158

\bibitem[Garrington et al.(1988)]{Gar1988} Garrington, S.~T.,
Leahy, J.~P., Conway, R.~G., \& Laing, R.~A.\ 1988, \nat, 331, 147

\bibitem[Garrington et al.(1991)]{Gar1991} Garrington, S.~T.,
Conway, R.~G., \& Leahy, J.~P.\ 1991, \mnras, 250, 171

\bibitem[Gawro{\'n}ski et al.(2006)]{Gawr2006} Gawro{\'n}ski, M.~P., 
Marecki, A., Kunert-Bajraszewska, M., \& Kus, A.~J.\ 2006, \aap, 447, 63

\bibitem[Gilbert et al.(2004)]{Gil2004} Gilbert, G.~M., Riley, J.~M., 
Hardcastle, M.~J., et al.\ 2004, \mnras, 351, 845 (G04)

\bibitem[Gopal-Krishna et al.(1996)]{GWH1996} Gopal-Krishna,
Wiita, P.~J., \& Hooda, J.~S.\ 1996, \aap, 316, L13

\bibitem[Gopal-Krishna \& Wiita(2000)]{GKW2000}
Gopal-Krishna, \& Wiita, P.~J.\ 2000, \aap, 363, 507

\bibitem[Hardcastle et al.(1997)]{Har1997} Hardcastle, M.~J., Alexander, P., 
Pooley, G.~G., \& Riley, J.~M.\ 1997, \mnras, 288, 859

\bibitem[Harvanek \& Hardcastle(1998)]{HH1998}
Harvanek, M., \& Hardcastle, M.~J.\ 1998, \apjs, 119, 25

\bibitem[Jorstad \& Marscher(2008)]{JM2008} Jorstad, S.~G., \&
Marscher, A.~P.\ 2008, Extragalactic Jets: Theory and Observation from Radio
to Gamma Ray, ASP Conference Series 386, 219

\bibitem[Kaiser et al.(2000)]{KSR2000} Kaiser, C.~R., Schoenmakers, A.~P., 
R\"ottgering, H.~J.~A.\ 2000, \mnras, 315, 381

\bibitem[Kharb et al.(2008)]{Kha2008} Kharb, P., O'Dea, C.~P., Baum, S.~A., 
et al.\ 2008, \apjs, 174, 74

\bibitem[Konar et al.(2012)]{Kon2012} Konar, C., Hardcastle, M.~J.,
Jamrozy, M., Croston, J.~H., \& Nandi, S.\ 2012, \mnras, 424, 1061

\bibitem[Laing(1988)]{Lai1988} Laing, R.~A.\ 1988, \nat, 331, 149

\bibitem[Law-Green et al.(1995)]{Law1995} Law-Green, J.~D.~B., Leahy, J.~P., 
Alexander, P., et al.\ 1995, \mnras, 274, 939

\bibitem[Longair \& Riley(1979)]{LR1979}
Longair, M.~S., \& Riley, J.~M.\ 1979, \mnras, 188, 625

\bibitem[Lonsdale \& Morison(1983)]{LM1983}
Lonsdale, C.~J., \& Morison, I.\ 1983, \mnras, 203, 833

\bibitem[Lonsdale et al.(1993)]{Lon1993} Lonsdale, C.~J.,
Barthel, P.~D., \& Miley, G.~K.\ 1993, \apjs, 87, 63

\bibitem[Marecki \& Swoboda(2011)]{MS2011} Marecki, A.,
\& Swoboda, B.\ 2011, \aap, 525, A6

\bibitem[Marecki(2012)]{Mar2012} Marecki, A.\ 2012, \aap, 544, L2 (Paper~I)

\bibitem[Neff et al.(1995)]{Nef1995} Neff, S.~G.,
Roberts, L., \& Hutchings, J.~B.\ 1995, \apjs, 99, 349

\bibitem[Pirya et al.(2011)]{Pir2011} Pirya, A., Nandi, S.,
Saikia, D.~J., \& Singh, M.\ 2011,\\
Bulletin of the Astronomical Society of India, 39, 547 {\tt arXiv:1201.0922}

\bibitem[Price et al.(1993)]{Pri1993} Price, R., Gower, A.~C.,
Hutchings, J.~B., et al.\ 1993, \apjs, 86, 365

\bibitem[Reid et al.(1995)]{Rei1995} Reid, A., Shone, D.~L., Akujor, C.~E., 
et al.\ 1995, \aaps, 110, 213

\bibitem[Reid et al.(1999)]{Rei1999} Reid, R.~I.,
Kronberg, P.~P., \& Perley, R.~A.\ 1999, \apjs, 124, 285

\bibitem[Riley et al.(1999)]{Ril1999} Riley, J.~M., Rawlings, S.,
McMahon, R.~G., et al.\ 1999, \mnras, 307, 293

\bibitem[Saikia \& Jamrozy(2009)]{SJ2009} Saikia, D.~J., \& Jamrozy, M.\ 2009,\\
Bulletin of the Astronomical Society of India, 37, 63 {\tt arXiv:1002.1841}

\bibitem[Schoenmakers et al.(2000)]{Sch2000} Schoenmakers, A.~P.,
de Bruyn, A.~G., R\"ottgering, H.~J.~A., \& van der Laan, H.\ 2000,
\mnras, 315, 395

\end{thebibliography}
\end{document}